\documentclass[aps, 10pt, notitlepage, twocolumn, superscriptaddress,
nofootinbib,longbibliography]{revtex4-2}

\usepackage{amsmath}
\usepackage{amssymb}
\usepackage{amsfonts}
\usepackage[utf8]{inputenc}
\usepackage[T1]{fontenc}
\usepackage{mathrsfs}
\usepackage{anyfontsize}

\usepackage[linktocpage,breaklinks]{hyperref}
\usepackage[usenames,dvipsnames]{xcolor}

\usepackage{txfonts}
\usepackage{tensor}
\usepackage{bm}

\usepackage{graphicx}
\usepackage{epsfig}
\usepackage{epstopdf}

\usepackage{natbib}

\usepackage{url}  
\usepackage{hyperref}

\hypersetup{colorlinks=true, citecolor=Purple, linkcolor=Purple,
urlcolor=Purple}

\usepackage{tikz}

\begin{document}

\title{Oscillating profiles from dark matter scalar solitons}
\author{Caio F. B. Macedo} \email{caiomacedo@ufpa.br}
\affiliation{Faculdade de F\'isica, Campus Salin\'opolis, Universidade Federal do Par\'a, 68721-000, Salin\'opolis, Par\'a, Brazil}
\author{Nicolas Aimar} \email{ndaimar@fe.up.pt}
\affiliation{Faculdade de Engenharia, Universidade do Porto,
s/n, R. Dr. Roberto Frias, 4200-465 Porto, Portugal}
\affiliation{CENTRA, Departamento de Fésica, Instituto Superior Técnico-IST,
Universidade de Lisboa-UL, Avenida Rovisco Pais 1, 1049-001 Lisboa, Portugal}
\author{Jo\~{a}o Lu\'{i}s Rosa} \email{joaoluis92@gmail.com}
\affiliation{Departamento de F\'isica Te\'orica and IPARCOS,
	Universidad Complutense de Madrid, E-28040 Madrid, Spain}
    \affiliation{Institute of Physics, University of Tartu, W. Ostwaldi 1, 50411 Tartu, Estonia}
\author{Diego Rubiera-Garcia} \email{drubiera@ucm.es}
\affiliation{Departamento de F\'isica Te\'orica and IPARCOS,
	Universidad Complutense de Madrid, E-28040 Madrid, Spain}

\begin{abstract}
Real scalar fields, e.g. the axion, cannot condensate into stationary solitonic configurations to form starlike structures, eventually either dispersing or collapsing. However, by relaxing the stationarity condition on the metric, it has been shown that oscillatory solitonic solutions---known as oscillatons---exist. Oscillatons share several properties with boson stars, including comparable compactness and mass ranges. However, their time-dependent nature can lead to potentially discriminating observable signatures.
In this work, we explore the observational properties of oscillatons. We find that stable oscillatory circular orbits exist, extending down to the center of the configuration, supporting the possibility of accretion disk structures within the star. We compute the deflection of light rays and verify that it is largely insensitive to the time dependence of the metric. Despite this, the oscillatory behavior of the redshift factor has a strong effect on the observed intensity profiles from accretion disks, producing a breathinglike image whose frequency depends on the mass of the scalar field. In fact, their oscillation period may lie within the observational windows of the Event Horizon Telescope for Sgr~A$^{*}$ and M87$^{*}$,
suggesting that this ``twinkling'' behavior may provide a potential observable signature of time-dependent compact objects. A detailed assessment of detectability in realistic interferometric observations is left for future work.
\end{abstract}

\maketitle

%%%%%%%%%%%%%%%%%%%%%%%%%%%%%%%%%%%%%%%%%%%%%%%%%%%%%%%%%%%%%%%%%%%%%%

\section{Introduction}
\label{sec:introduction}

Although scarce within the Standard Model (SM), self-interacting scalar and pseudoscalar fields arise naturally in many extensions of the SM. A prominent example is the quantum chromodynamics (QCD) axion, originally proposed to solve the strong charge-parity (CP) problem by promoting the CP-violating angle to a dynamical field~\cite{PhysRevLett.38.1440,Peccei:2006as}. At low energies, the QCD axion can be effectively treated as a scalar field with a self-interacting potential whose structure is dictated by nonperturbative QCD effects and depends on a model-dependent decay constant. 
Beyond the QCD construction, string-theoretic compactifications generically predict a large number of ultralight pseudoscalar degrees of freedom---the so-called \emph{string axiverse}---whose masses may span many orders of magnitude~\cite{Arvanitaki:2009fg}. This spectrum enables a variety of potentially observable phenomena, including superradiant instabilities of rotating black holes \cite{Brito:2015oca}, the formation of gravitational ``atoms'' \cite{Arvanitaki:2014wva,Bah:2025vbr}, modifications to extreme mass-ratio inspiral (EMRI) dynamics~\cite{Duque:2023seg}, constraints from black hole spin measurements~\cite{Ng:2020ruv}, and viable dark-matter candidates~\cite{Ferreira:2020fam}. The confluence of these signatures makes axionlike fields an attractive and well-motivated sector for beyond-SM physics.

Another intriguing consequence of ultralight bosonic fields is their ability to form compact, self-gravitating configurations through gravitational condensation (see the reviews~\cite{Schunck:2003kk,Liebling:2012fv} for details). These ``starlike'' objects appear in many forms, depending on the field content, self-interaction potential, charges, and possible couplings to additional matter sectors. For complex scalar fields, stationary solutions arise from imposing a harmonic time-dependence, yielding boson stars supported by the balance between gravity and the dispersive nature of the field. In the minimally coupled massive case, the maximum mass of the ground-state configuration is $M_{\rm max} \approx  0.633\, M_{\rm Pl}^2/\mu$~\cite{Kaup:1968zz,Ruffini:1969qy}, where $M_{\rm Pl}$ is Planck's mass, and $\mu$ the scalar field mass. Depending on the latter, such stars may span a wide range of astrophysical scales \cite{Liebling:2012fv}, may be revealed through gravitational-wave observations \cite{Palenzuela:2017kcg,Hannuksela:2018izj,Macedo:2013jja,Duque:2023seg}  and, due to their extremely weak interactions with SM fields, they can serve as compelling proxies for dark compact objects---for instance as alternatives to supermassive black holes at galactic centers \cite{Vincent:2015xta,Olivares:2018abq} or as dark-matter substructures \cite{Hui:2016ltb}.

{Compact bosonic solitons are not restricted to minimally coupled scalar fields.
Indeed, similar self-gravitating configurations arise in a broad variety of
bosonic theories wDith different spin content and interaction structures. For a
general review on boson stars and oscillatons, see Ref.~\cite{Visinelli:2021uve}. Vector
bosonic fields, for instance, can form compact objects known as
Proca stars~\cite{Brito:2015pxa}, which share several structural and dynamical
properties with scalar boson stars. More generally, nonrelativistic solitonic
configurations sourced by spin-0, spin-1, and spin-2 bosonic fields have also
been investigated in unified frameworks~\cite{Jain:2021pnk}, including recent
studies of spin-2 gravitational solitons~\cite{Schiappacasse:2025mao}. These
results suggest that many qualitative features of bosonic compact objects may
extend beyond the scalar case, motivating a broader phenomenological study of
self-gravitating bosonic configurations.}

Importantly, many features of complex-scalar {and vector} boson stars extend naturally to other types of bosonic fields. Real scalar fields, for instance, do not admit static, everywhere regular solitonic solutions \cite{Derrick:1964ww,Khlopov:1985fch}, but they \emph{do} form long-lived, localized, time-periodic configurations known as \emph{oscillatons}~\cite{Seidel:1991zh}. These solutions arise from the fully nonlinear interplay between gravity and the scalar field, and their existence broadens the range of possible compact objects seeded by light fields beyond the SM. Interestingly, the maximum mass configuration of oscillatons is quite similar to their complex cousins, given by $M_{\rm max} \approx 0.604\, M_{\rm Pl}^2/\mu$. However, due to their time-dependent behavior in the geometry, they could lead to different signatures as far as observations are concerned (see, e.g., Refs.~\cite{Ferreira:2017pth,Boskovic:2018rub,Ferreira:2019hzf,GRAVITY:2019tuf,GRAVITY:2023cjt,Kim:2024rgf}). Of particular interest for this work are those that involve the motion of light rays about oscillatons in view of the possibilities brought by current and future observational devices within this context. 

The recent horizon-scale observations by the Event Horizon Telescope (EHT) of M87$^*$ \cite{Akiyama:2019eap} and Sgr~A$^*$ \cite{Akiyama:2022l12} supermassive central objects have opened an unprecedented window into the strong-gravity regime, providing direct measurements of the size and morphology of the emitting region surrounding supermassive compact objects~\cite{EventHorizonTelescope:2019dse,EventHorizonTelescope:2022wkp}. These observations are broadly consistent with the shadow predicted by general relativity for a Kerr black hole \cite{Falcke:1999pj,Gralla:2019xty,Chael:2021rjo,Vincent:2022fwj}, yet the current disk-modeling and observational uncertainties still allow for the possibility of alternative compact configurations \cite{Vagnozzi:2022moj}, in particular those that lack an event horizon. Among the most theoretically motivated candidates are stars supported by ultralight bosonic fields, including boson stars, oscillatons, and more general solitonic condensates. As emphasized in Refs.~\cite{Rosa:2022tfv,Rosa:2023hfm,Rosa:2022toh,Rosa:2023qcv,Rosa:2024bqv,Rosa:2024eva,Tamm:2023wvn,Rosa:2025dzq}, horizon-scale imaging places important constraints on the allowed compactness, light-bending properties, and emission profiles of these configurations, thereby linking EHT measurements to fundamental physics questions regarding the existence and role of ultralight fields in astrophysics and cosmology. {Moreover, the next-generation EHT is expected to significantly improve angular resolution, dynamic range, temporal coverage, and multifrequency imaging capabilities, opening the prospect of probing smaller-scale and time-dependent features in horizon-scale images~\cite{Ayzenberg:2023hfw}.}

In this work, we investigate in detail the observational signatures of oscillatons—time-dependent,
self-gravitating configurations of real scalar fields—as potential alternatives to canonical black
holes and complex-scalar boson stars. We begin by constructing the background oscillaton
solutions using a Fourier-expanded Einstein--Klein--Gordon system and identify the most compact,
physically relevant configuration for our analysis. We then perform a comprehensive study of
geodesic motion, showing that oscillatory circular orbits (OCOs) exist at all radii, including deep
within the stellar interior, and that their oscillatory and epicyclic structure closely parallels the
stable circular motions around static boson stars. Building on these results, we analyze null
geodesics and light bending, demonstrating that oscillaton spacetimes generically produce caustics
and rainbow-type scattering (see also Ref.~\cite{Yang:2025yej}).
We then compute accretion-disk emission profiles, adopting both central-emission and
Novikov-Thorne-type models, and show that the intrinsic time dependence of the metric imprints a
distinctive ``breathing'' pattern on the observed intensity. This effect produces periodic transitions
between central brightening and ringlike morphologies, with a frequency set by the scalar-field
mass. Finally, we assess the detectability of these oscillations for supermassive objects and argue
that, for the mass scales relevant to galactic centers, the Event Horizon Telescope could in
principle resolve such ``twinkling'' features, providing a unique observational window into
ultralight-field solitons.

The remainder of this paper is organized as follows. In Sec.~\ref{sec:framework} we lay out the basic equations to be solved, show the specific \textit{ansatz} and the solution to be used as a proxy for the rest of the paper. In Sec.\ref{sec:orbitstime} we analyze the geodesic structure of the oscillaton spacetimes and the properties of oscillating circular bound orbits. In Sec. \ref{sec:disks} we introduce two models for the emission profile of the accretion disks, one following the Novikov-Thorne model and another considering an emission from the interior of the oscillaton caused by matter accumulation, and we produce the respective observed intensity profiles and shadow images. Finally, in Sec. \ref{sec:conclusion}, we trace our conclusions and prospects for future work. In the remainder of this paper we use natural units ($G=c=\hbar=1$)

\section{Theoretical framework and background} \label{sec:framework}

Oscillatons are solutions of Einstein’s equations minimally coupled to a real scalar field. The action reads\footnote{Note that we have adopted a system of geometrized units following which $G=c=1$, where $G$ is the gravitational constant and $c$ is the speed of light, such that $\frac{8\pi G}{c^4} = 8\pi$.}
\begin{equation}\label{eq:action}
    S = \int_\mathcal{M} \sqrt{-g}\left[\frac{R}{16\pi}
    -\frac{1}{2}\nabla_a\Phi\nabla^a\Phi
    -\frac{1}{2}V(\Phi)\right]d^4x,
\end{equation}
where $R$ is the Ricci scalar, $g$ the determinant of the space-time metric $g_{ab}$, $\Phi$ a real-valued scalar field, $V(\Phi)$ is the self-interaction potential of the scalar field, $\nabla_a$ denotes a covariant derivative, and $x^a$ represents the coordinate system on the spacetime manifold $\mathcal{M}$. In this work, we focus on the simplest potential that supports spherically symmetric solutions, namely
\[
    V(\Phi) = \mu^2 \Phi^2,
\]
where $\mu$ is a parameter that plays the role of the mass of the scalar field. Taking a variation of Eq. \eqref{eq:action} with respect to $g_{ab}$ and $\Phi$, respectively, yields the Einstein–Klein–Gordon system
\begin{align}
    G_{ab} &= 8\pi T^{(\Phi)}_{ab},\label{eq:einstein}\\
    \nabla_a\nabla^a\Phi &= \frac{dV}{d\Phi},\label{eq:phi}
\end{align}
where the stress–energy tensor of the scalar field is
\begin{equation}
    T^{(\Phi)}_{ab} = \nabla_a\Phi\nabla_b\Phi
    -\frac{1}{2}g_{ab}\left[\nabla_c\Phi\nabla^c\Phi+V(\Phi)\right].
\end{equation}

Static and spherically symmetric solutions of Eqs.~\eqref{eq:einstein}–\eqref{eq:phi} do not exist: any initial scalar-field configuration either disperses or collapses~\cite{Reiris2015,Heusler1996,Bekenstein1995}. One way to circumvent this is to allow both the scalar field and the metric fields to be time-dependent. Although these solutions are not truly stable since they eventually disperse, their decay timescale is extremely long, often exceeding the age of the Universe for physically relevant cases~\cite{Page2003,Fodor2010,Zhang2020}. We therefore consider the ansatz
\begin{align}
    \Phi &= \Phi(t,r),\label{eq:phi_g}\\
    ds^2 &= -A(t,r)\,dt^2 + B(t,r)^{-1}dr^2 + r^2 d\Omega^2,\label{eq:metric_g}
\end{align}
where $x^a=(t,r,\theta,\varphi)$ are standard Schwarzschild spherical coordinates, $A$ and $B$ are metric functions, and $d\Omega^2=d \theta^2 + \sin^2 \theta d \varphi^2$ is the line element of the unit 2-sphere. Solutions of this form are known as \textit{oscillatons}~\cite{Seidel:1991zh}.

Substituting Eqs. \eqref{eq:phi_g}–\eqref{eq:metric_g} into Eqs.~\eqref{eq:einstein}–\eqref{eq:phi}, one obtains
\begin{align}
    \frac{1-r B'-B}{r^2} &= 4\pi\!\left(\mu^2 \Phi^2 + \frac{\dot{\Phi}^2}{A} + B\,\Phi'^2\right),\label{eq:osc1}\\
    \frac{1-\left(\tfrac{r A'}{A}+1\right) B}{r^2} &= 4\pi\!\left(\mu^2 \Phi^2 - \frac{\dot{\Phi}^2}{A} - B\,\Phi'^2\right),\\
    A B \Phi'' - \ddot{\Phi} 
    &+ \tfrac{1}{2}\left(\frac{\dot{A}}{A}\dot{\Phi} - 2\mu^2 A \Phi + B \dot{B}\,\dot{\Phi}\right)\nonumber\\
    &+ \frac{\Phi'\!\left(r A' B + A(r B' + 4B)\right)}{2r} = 0.\label{eq:osc3}
\end{align}
The above equations cannot be easily solved by standard methods. However, as noted in Ref.~\cite{Seidel:1991zh}, one can expand the metric functions and the scalar field in a discrete Fourier series with frequency $\omega$, namely
\begin{align}
    A(t,r) &= \sum_{j=0}^{N} \tilde{A}_j(r)\cos(2j\omega t),\label{eq:A}\\
    B(t,r) &= \sum_{j=0}^{N} \tilde{B}_j(r)\cos(2j\omega t),\label{eq:B}\\
    \Phi(t,r) &= \sum_{j=0}^{N} \tilde{\phi}_j(r)\cos\!\big[(2j+1)\omega t\big],\label{eq:phie}
\end{align}
where $N$ is chosen to be large enough to guarantee the convergence of the summation. In practice, we consider terms up to $N=3$, which are sufficient, as we illustrate below.

Substituting Eqs.~\eqref{eq:A}–\eqref{eq:phie} into Eqs.~\eqref{eq:osc1}–\eqref{eq:osc3} and exploiting the orthogonality of the Fourier functions, one obtains a coupled system of ordinary differential equations (in $r$) for the functions $\tilde{A}_j(r)$, $\tilde{B}_j(r)$, and $\tilde{\phi}_j(r)$. 
To close the system, appropriate boundary conditions are required. For oscillatons, one demands regularity at the center and asymptotic flatness at infinity. Regularity at the center requires $\tilde\phi_j'(0)=0 $, $\tilde B_0=1$ and $\tilde B_j=0$, for $j>1$. Asymptotic flatness requires $\tilde A_j(r\to\infty)=\tilde B_j(r\to \infty)$, and $\tilde \phi_j(r\to\infty)=0$. Given a central value of the scalar field term $\tilde{\phi}_0(0)=\phi_c$, this becomes a boundary-value problem for the parameters $\tilde A_j(0)$, $\tilde{\phi}_j(0)$, and $\omega$. We solve the above system using a shooting method, imposing boundary conditions at the origin and adjusting the parameters to enforce asymptotic flatness. At spatial infinity, we extract the total mass of the star, defined as
\begin{equation}
    M = m(r\to \infty) 
    = \left.\tfrac{1}{2}\,[1-B_0(r)]\,r\right|_{r\to\infty}.
\end{equation}
Note that the calculation above considers only the \textit{time-independent} part of the metric, which is equivalent to averaging the metric functions over time. We also define an effective radius $R$ of the oscillaton as the radius that encapsulates {$98\%$ of the total mass, i.e., $m\left(R\right)=0.98M$.}

\begin{figure}[t!]
    \centering
    \includegraphics[width=\linewidth]{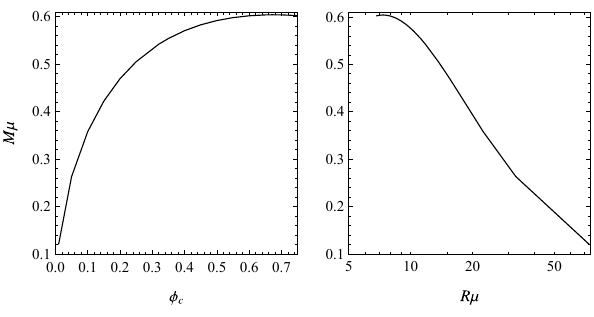}
    \caption{Sequence of oscillaton solutions. \textit{Left panel}: total mass $\mu M$ as a function of $\phi_c$. 
    \textit{Right panel}: mass–radius relation for the oscillaton configurations. We can see that oscillaton shares many similarities with standard boson stars (see also Fig. 1 in Ref.~\cite{Brito:2015yfh}).}
    \label{fig:config}
\end{figure}

The above procedure generates a one-parameter family for each value of $\phi_c$, considering node-less solutions. In the left panel of Fig.~\ref{fig:config}, we show the total mass of the oscillaton as a function of  $\phi_c$. Oscillatons closely resemble boson stars: they feature a maximum mass, in this case $\mu M \approx 0.604$, which is reached at a critical central field $\phi_c \approx 0.67$. The maximum mass is close to the Kaup limit for the boson star, which is $\mu M\approx0.633$~\cite{Kaup:1968zz}. The right panel of Fig.~\ref{fig:config} displays the corresponding mass–radius relation. As we mentioned above, although oscillatons are generically unstable, configurations to the left of the maximum mass have been shown to exhibit a distinct type of instability with a timescale much shorter than the usual scalar-field leakage. For our analysis, we therefore focus on the most compact viable configuration, characterized by $(\phi_c,\omega/\mu,M\mu)=(0.67,0.864,0.604)$. Nonetheless, we highlight that the orbital features presented here are universal for less compact configurations, the major difference being the redshift intensity.

\begin{figure*}
    \centering
    \includegraphics[width=\linewidth]{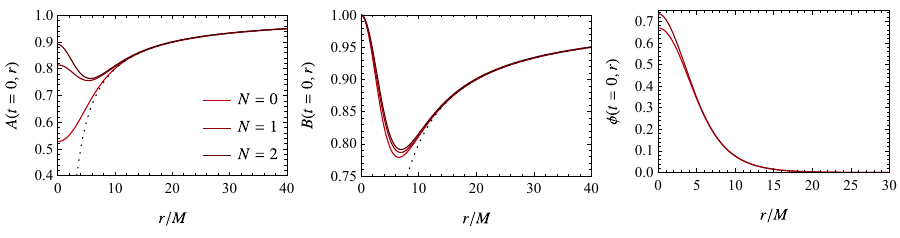}
    \caption{Background metric and scalar field for the maximum-mass oscillaton. 
    Darker shades of red correspond to including more terms in the series expansions 
    \eqref{eq:A}–\eqref{eq:phie} (up to $N=2$). The dotted line indicates the corresponding Schwarzschild solution.}
    \label{fig:background}
\end{figure*}

To illustrate the profiles for the solutions as well as the convergence of the series in Eqs. \eqref{eq:A}--\eqref{eq:phie},  in Fig.~\ref{fig:background} we show the background metric and scalar field for the configuration we adopt, which is also representative of typical oscillaton profiles (see, e.g., Refs.~\cite{Seidel:1991zh,Urena-Lopez:2001zjo,Alcubierre:2003sx,Brito:2015yga}). We can see how additional terms contribute to the expansion in the configurations. As before, we truncate the Fourier expansions at $N=2$, which is sufficient for our purposes. For this configuration, the radius of the star is given by $R\mu=7.376$ ($R/M=12.21$).
{We notice that similar expansions were used in the literature, including in the seminal work that introduced the oscillaton solution~\cite{Seidel:1991zh} and in vector extension of the oscillatons~\cite{Brito:2015yfh}. }

{We emphasize that the oscillaton solutions considered here are based on the
well-established Fourier-expansion approach originally introduced in
Ref.~\cite{Seidel:1991zh} and later explored in several contexts, including vector oscillatons
and related compact solitonic configurations. We have independently
cross-checked our numerical solutions against the results available in the
literature, finding excellent agreement for the background profiles, maximum
mass, and effective radius. Figure~\ref{fig:background} also illustrates the convergence of the
Fourier expansion for the most compact configuration considered in this work,
showing that the inclusion of higher harmonics produces only small corrections
to the metric and scalar-field profiles. Since the main goal of this work is to
investigate the phenomenology associated with the oscillatory background,
particularly the modulation of the redshift factor, we restrict the analysis to
the truncation order required for quantitative convergence of the observables
considered here.}

{We notice here that the oscillaton field decays exponentially due to the mass term but it is never exactly zero. Therefore, the spacetime only approaches Schwarzschild as $r\to\infty$. However, in practice, one matches the spacetime with the Schwarzschild one at a given outer radius, which in our case is $r \mu\approx35$. 
Since the oscillaton spacetime is asymptotically flat, the use of a Schwarzschild approximation at large radii is purely a numerical convenience to improve integration efficiency. We have verified that varying the matching radius does not affect the resulting deflection angles within numerical accuracy.
}

\section{Orbital motion, oscillatory timelike circular geodesics and light-rays}\label{sec:orbitstime}

\begin{figure*}
    \centering
    \includegraphics[width=1\linewidth]{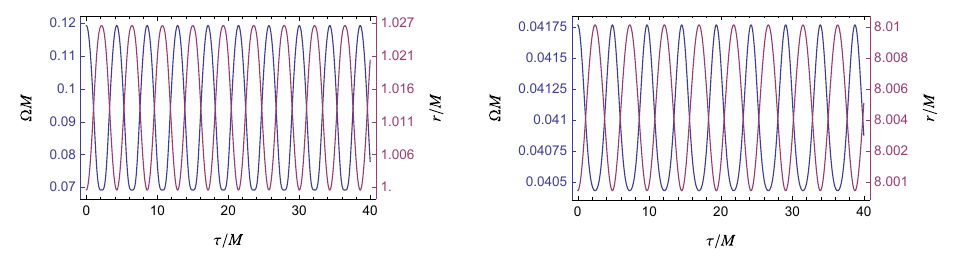}
    \caption{Oscillatory patterns in the orbital frequency and radius of OCOs for $(r_0,L)=(1,0.12834)$ (left panel) and $(r_0,L)=(8,3.262266)$ (right panel).}
    \label{fig:circular}
\end{figure*}

Unlike the case of static boson stars, circular orbits cannot be strictly defined in oscillaton spacetimes. This is a direct consequence of the time dependence of the metric functions. However, since the oscillatory terms are typically smaller than the time-independent part, one can still define \textit{oscillatory circular orbits} (OCOs), i.e., bound orbits along which the orbital radius oscillates around a constant value.
{We define an OCO as a bound trajectory whose radial coordinate remains confined around a mean radius, with the angular momentum chosen numerically so as to minimize the radial excursion over an oscillation period of the background. These trajectories are therefore not circular orbits in the strict stationary-spacetime sense, but rather the natural time-periodic generalization of circular motion in an oscillating geometry.}
In what follows, we investigate the existence of OCOs in oscillaton spacetimes, and show how they reduce to the usual circular Schwarzschild orbits in the appropriate limit.

Geodesic motion can be analyzed through the one-particle Lagrangian density
\begin{equation} 
   2\mathcal{L} = -A\,\dot{t}^2 + B\,\dot{r}^2 + r^2\dot\varphi^2 = -\delta,\label{eq:const}
\end{equation}
where dots denote derivatives with respect to the affine parameter $\lambda$ (which is the proper time $\tau$ for timelike geodesics), and $\delta=1$ $(\delta=0)$ for timelike (null) geodesics. From Eq.~\eqref{eq:const}, we see that the specific angular momentum remains conserved:
\begin{equation}\label{eq:angL}
    L = \frac{\partial \mathcal{L}}{\partial\dot\varphi} = r^2\dot\varphi.
\end{equation}
However, because of the time dependence of the metric, the specific energy is not conserved. Instead, one must obtain a system of differential equations for $(t,r,\varphi)$ and perform an integration of them. From the Euler–Lagrange equations,
\begin{equation}
    \frac{\partial\mathcal{L}}{\partial x^a}
    - \frac{d}{d\lambda}\left(\frac{\partial\mathcal{L}}{\partial \dot x^a}\right)=0,
\end{equation}
we obtain, from Eq.~\eqref{eq:const}, for the coordinates $t$ and $r$
\begin{align}
    \ddot t 
    &+ \frac{\partial_r A}{A}\,\dot r \dot t
    + \frac{\partial_t A}{2A}\,\dot t^2
    - \frac{\partial_t B}{2AB^2}\,\dot r^2 = 0,\label{eq:ddt}\\
    \ddot r 
    &+ \tfrac{1}{2}\,\partial_r A\,B\,\dot t^2
    - \frac{\partial_t B}{B}\,\dot r \dot t
    - \frac{\partial_r B}{2B}\,\dot r^2
    - \frac{L^2 B}{r^3} = 0.\label{eq:ddr}
\end{align}

A useful quantity to analyze the orbital motion is the angular frequency measured by an observer at infinity, defined as
\begin{equation} \label{eq:angvec}
    \Omega = \frac{\dot\varphi}{\dot t}
           = \frac{L}{r^2\dot t},
\end{equation}
where we have used Eq. \eqref{eq:angL} to eliminate $\dot\varphi$ in terms of $L$. As $\dot t$ is a numerical solution, we cannot explicitly write it in terms of the orbital-specific energy for circular orbits. For comparison, in static spacetimes, considering circular orbits, one finds
\[
    \Omega = \sqrt{\frac{A'(r)}{2r}},
\]
which, in the Schwarzschild case, reduces to the Keplerian frequency,
\[
    \Omega_{\rm K} = \sqrt{\frac{M}{r^3}}.
\]

Once a value for the specific angular momentum $L$ is set, Eqs.~\eqref{eq:angL}, \eqref{eq:ddt}, and \eqref{eq:ddr} can be integrated, while the constraint in Eq. \eqref{eq:const} provides a relation among the initial conditions. To investigate OCOs, we integrate the system under the following initial conditions:
\begin{equation}
    r(0)=r_0,\quad \dot r(0)=0,\quad t(0)=0,\quad \varphi(0)=0,
\end{equation}
while $\dot t(0)$ is determined from Eq.~\eqref{eq:const}. We then numerically search for orbital motions along which the radial coordinate exhibits minimal variation.  

\begin{figure}
    \centering
    \includegraphics[width=1\linewidth]{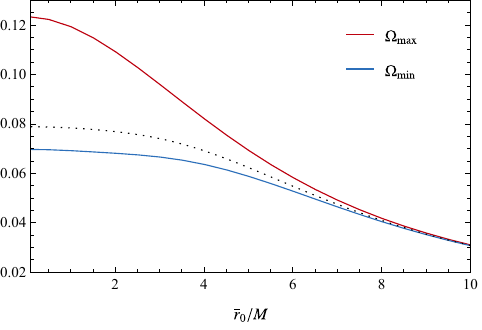}
    \caption{Minimum and maximum values of the angular frequency in Eq. (\ref{eq:angvec}) as a function of the average orbital radius of the OCO. The dotted black line corresponds to $(\tilde A_{0}'(r)/2r)^{1/2}$, which asymptotically approaches the Keplerian frequency.}
    \label{fig:frequency}
\end{figure}

Examples of OCOs are shown in Fig.~\ref{fig:circular}, showing the cases $(r_0,L)=(1,0.12834)$ and $(r_0,L)=(8,3.262266)$. These values were chosen to illustrate two distinct regimes: one very close to the central oscillations of the star and another near the surface, though still within the stellar interior. In both cases, the orbits are qualitatively similar: the radial position and angular frequencies oscillate with the same frequency but in the opposite phase. The frequency of the oscillatory pattern is multiples of $2\omega$, as is natural to expect due to the behavior of the metric in Eqs. \eqref{eq:A} and \eqref{eq:B}\footnote{In our case, we see only $2\omega$ and $4\omega$ in the spectral analysis of the OCO, due to the truncation in $N$.}. Moreover, the closer the orbit lies to the stellar center, the more pronounced the difference between the maximum and minimum angular frequencies becomes, although the radial variation remains small. This behavior is quantified in Fig.~\ref{fig:frequency}, which shows the minimum and maximum values of the angular frequency as a function of the average orbital radius $\bar r_0$\footnote{Since the difference between the maximum and minimum radii is generally small, we may approximate $r_{\rm min}\approx r_{\rm max}\approx \bar r_0$.}. As expected, the discrepancy between the maximum and minimum values decreases with $\bar r_0$, reflecting the decay of the amplitude of the metric oscillations with the distance from the stellar center. Asymptotically, these values approach $(\tilde A_0'(r)/2r)^{1/2}$, shown as the black dotted line in Fig.~\ref{fig:frequency}.

\begin{figure}
    \centering
    \includegraphics[width=1\linewidth]{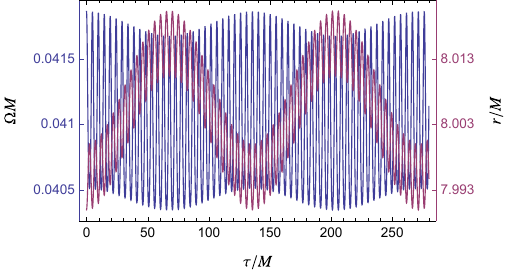}
    \caption{Slightly disturbed OCO orbit. For this orbital motion, we consider $(r_0,L)=(7.99,3.262266)$, so as to compare with the right-panel of Fig.~\ref{fig:circular}. We see that there is an additional oscillatory pattern, reminiscent from the epicyclic modulation.}
    \label{fig:epicyclic}
\end{figure}

\begin{figure}
    \centering
    \includegraphics[width=1\linewidth]{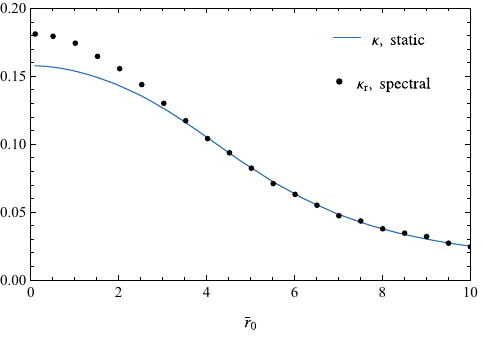}
    \caption{Epicyclic frequency computed through Eq.~\eqref{eq:epi} by using only the time-independent part of the metric, and the values extracted from a spectral analysis of slightly disturbed OCOs, as a function of $\bar r_0$. }
    \label{fig:spectral}
\end{figure}

It is natural to ask whether these OCOs are stable, i.e., whether a small perturbation in the parameters causes the orbiting particle to drift away from its trajectory. Figure~\ref{fig:epicyclic} shows one such case, with $(r_0,L)=(7.99,3.262266)$, similar to the orbit illustrated in the right panel of Fig.~\ref{fig:circular}. In addition to the usual oscillatory pattern of OCOs, an additional low-frequency modulation is visible. This resembles an epicyclic oscillation, where a perturbed stable orbit oscillates around its equilibrium configuration (see, e.g., Ref.~\cite{BinneyTremaine2008}).  
In static spacetimes, we find that the epicyclic frequency $\kappa$ of the radial motion is given by~\cite{DeFalco:2021btn}
\begin{equation}\label{eq:epi}
    \kappa^2=\frac{B\left(r A A''-2 r A'^2+3 A A'\right)}{2rA}\Bigg\vert_{r=r_0}.
\end{equation}
For the Schwarzschild spacetime this becomes
\[
\kappa^2=M\left(\frac{r_0-6M}{r_0^4}\right),
\]
which indicates that orbital motions with $r_0<6M$ are unstable. We find that OCOs, when perturbed, exhibit a similar epicyclic frequency even deep inside the stellar configurations. This suggests that OCOs are the natural generalization of stable circular orbits in oscillatons, existing for all orbital radii, in close analogy to the case of static scalar boson stars~\cite{Rosa:2023qcv}.

Although it is not straightforward to analytically study the radial ``epicyclic'' frequency of perturbed OCOs, we can perform a spectral analysis of orbital motions such as the one shown in Fig.~\ref{fig:epicyclic}. From the corresponding power spectrum, three prominent frequencies emerge: two at $2\omega$ and $4\omega$, originating from the background spacetime oscillations (therefore also being the OCO frequency), and a third, lower frequency $\kappa_{\rm r}$, corresponding to the epicyclic remnant. 
In Fig.~\ref{fig:spectral}, we show $\kappa_{\rm r}$ as a function of the average orbital radius $\bar r_0$, and compare it with the epicyclic frequency given by Eq.~\eqref{eq:epi}, calculated using only the stationary part of the metric. The agreement is excellent even for orbital motions as deep inside the stellar configuration as $r_0=4M$.

\subsection{Light-ray deflections and hot spot within oscillatons}

Let us now consider the case $\delta=0$, focusing on the propagation of null geodesics in oscillatons. We start with a stream of light rays emitted by an observer in the asymptotic region, far from the oscillaton. In this region, the spacetime is approximately Schwarzschild. As the rays approach the star, at some finite radius (typically taken to be four times the stellar radius), we switch to the oscillaton metric and continue the integration procedure. The relevant equations remain Eqs.~\eqref{eq:angL}, \eqref{eq:ddt}, and \eqref{eq:ddr}, together with the constraint in Eq.~\eqref{eq:const}.  

\begin{figure}
    \centering
    \includegraphics[width=\linewidth]{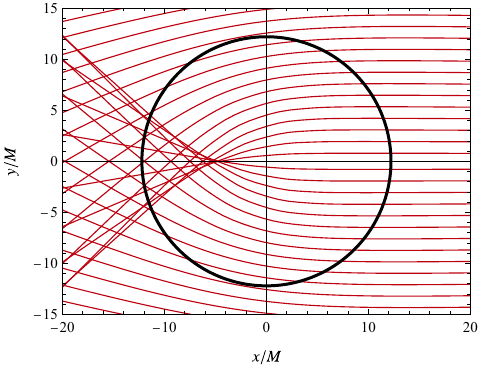}
    \caption{Light rays incoming from $r_0=800M$ toward the oscillation with different impact parameters, arriving almost parallel to the $x$-axis. We set $t(0)=0$ as the initial condition. The stellar radius is shown in black. A caustic forms with a cusp located at $r\approx 5M$.}
    \label{fig:deflection}
\end{figure}

\begin{figure}
    \centering
    \includegraphics[width=\linewidth]{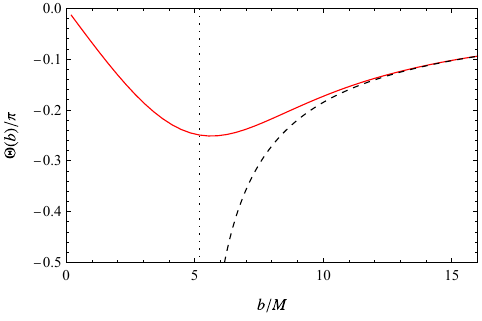}
    \caption{Deflection angle for the oscillaton configuration studied here, as a function of the impact parameter. The dashed line corresponds to the Schwarzschild black hole, which diverges at the critical impact parameter of this solution, corresponding to $b=3\sqrt{3}M$.}
    \label{fig:deflect_angle}
\end{figure}

The result is shown in Fig.~\ref{fig:deflection}, where we integrate with $r(0)=800M$, $\varphi(0)=0$, and $t(0)=0$, for different values of the angular momentum $L$ (or equivalently the impact parameter $b$). Unlike the static case, the light-ray trajectories depend on the initial time $t(0)$, since this determines the oscillation phase of the metric at emission. Nevertheless, our numerical results show that this dependence is very weak and the choice $t(0)=0$ is representative. We also notice the formation of a caustic, characterized by an extremum in the scattering angle, which in turn gives rise to rainbow scattering phenomenology. These features were already observed for compact uniform-density stars~\cite{Stratton:2019deq} and also in black holes surrounded by matter~\cite{Leite:2019uql}. In Fig.~\ref{fig:deflect_angle} we show the scattering angle as a function of the impact parameter.\footnote{Notice that we define the scattering angle in the same way as in Ref.~\cite{Stratton:2019deq}.} The precise rainbow angle can be extracted from the minimum of the scattering curve, which in our model yields $\theta_r \approx 45.2^{\circ}$. 

Unlike fluid stars, oscillatons are composed of scalar fields that interact with ordinary matter only through gravity. Consequently, the appearance of caustics and rainbow angles has a distinctive physical relevance: emitters may be located inside the star, including at the precise region where the cusp is formed. This opens up the intriguing possibility of detecting enhanced fluxes of light rays from sources -- such as hot spots -- placed at these locations. We emphasize, however, that this feature is not exclusive to oscillatons but also occurs in ordinary boson star configurations. The distinctive characteristic of oscillaton is that the time-dependent caustic could potentially generate observable effects, as recently explored in Ref.~\cite{Yang:2025yej}.

\begin{figure*}[t!]
    \includegraphics[width=.8\linewidth]{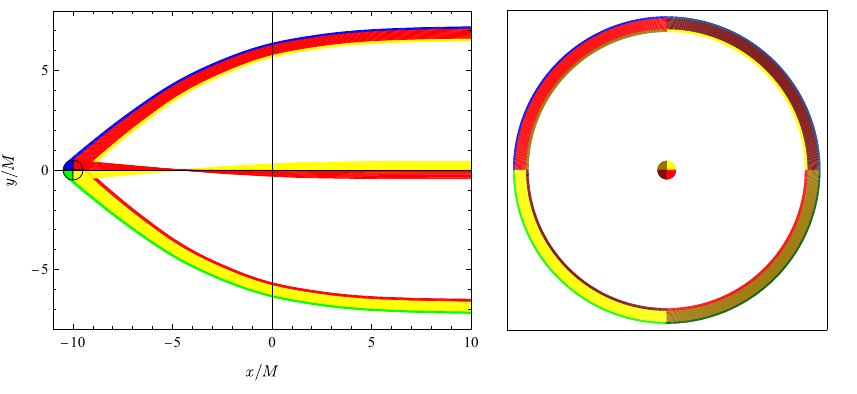}\\
    \includegraphics[width=.8\linewidth]{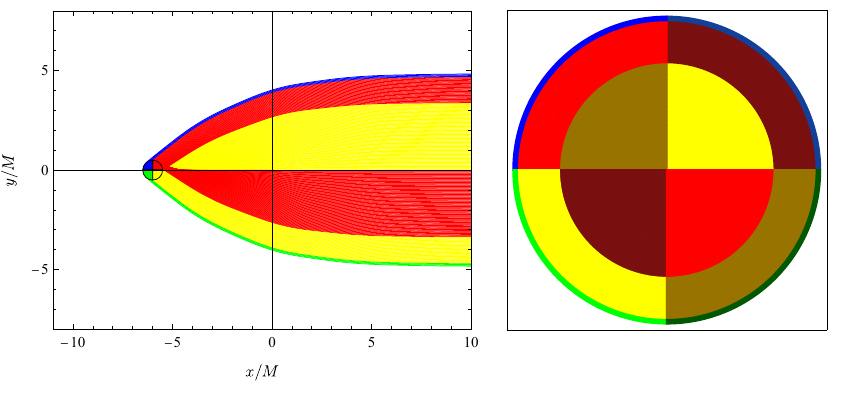}\\
    \includegraphics[width=.8\linewidth]{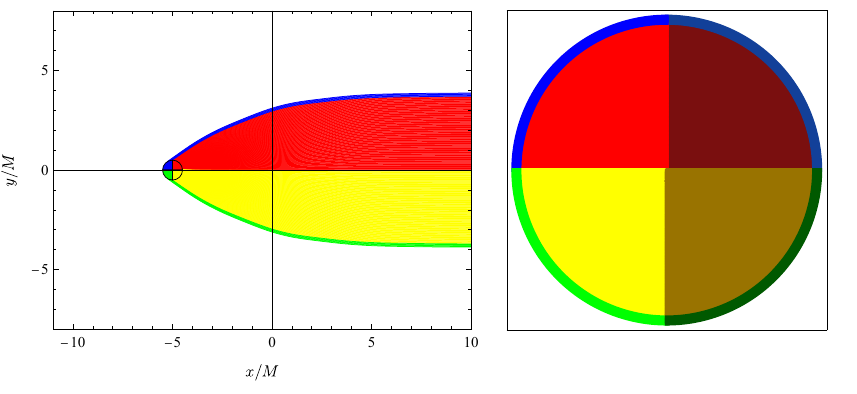}
    \caption{Light-rays emitted by a finite size object reaching an asymptotic observer. The radius of the emitter is set at $0.5M$ and its location is set at $-10M$ (upper panel), $-6M$ (lower panel), and $-5M$ (middle panel). In all cases, there is a region of the emitter which is not observed, although the back part of the sphere is partially seen. When the source is located at a radial position of $-10M$, we see the formation of the Einstein ring.}
    \label{fig:hotspots_def}
\end{figure*}

\begin{figure}[t!]
    \centering
    \includegraphics[width=1\linewidth]{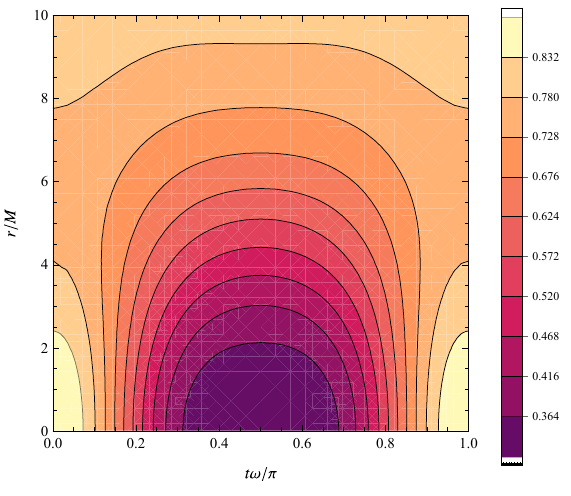}
    \caption{Metric $A(t,r)$ as function of time and radius. Matter distributions would be affected differently depending on their position at different instants of time. Notice that the maximum redshift location is displaced.}
    \label{fig:lapse}
\end{figure}
To illustrate how an object within the oscillaton would be seen, in Fig.~\ref{fig:hotspots_def} we analyze light rays emitted from the surface of a spherical object placed at $(x,y)=(-10M,0)$ (upper panel), $(x,y)=(-6M,0)$ (middle plane), and $(x,y)=(-5M,0)$, reaching an observer at $y=0$ for $x\gg M$. The color-code of the object is illustrated both in the representative circle (right panels of the figure) and the deflected rays. We represent the right hemisphere --- from the perspective of the observer --- in dark colors to better illustrate the lensing. Since the object's radius of $R_0=0.5M$ is small compared to the size of the star, we consider the redshift difference at different points on its surface to be negligible. In addition, we verify that the time delay between rays is very small. From Fig.~\ref{fig:hotspots_def} we see that, at some distance within the star, the lensing does not generate an Einstein ring, although the bending effect can still be strong enough to distort the images. We see three different regimes, one with the Einstein's ring (upper panel), one with a swapping in colors (middle panel), and a distorted image. Notice that even when there are no swaps, a small portion of the back of the object is visible due to the bending.

Apart from the bending of light, two additional effects occur in oscillatons that are absent in the usual boson star case. Due to the oscillatory behavior of the metric, any fixed light source will suffer \textit{i)} a redshift effect, and \textit{ii)} a change in the intensity profile, both as functions of time. The rate of change is related to the frequency $\omega$ of the background metric, according to the metric displayed in Eq.~\eqref{eq:A}. In Fig.~\ref{fig:lapse} we show the metric function at different locations of the star, as a function of time. We can see that the change in magnitude of $A(t,r)$ in time is more pronounced in regions deep within the star. As we have shown in Sec.~\ref{sec:orbitstime}, since stable timelike bound orbital motion exists inside the star, redshift changes of light emissions from accretion disks will be quite apparent from those regions. This is in clear contrast with the usual compact boson stars~\cite{Rosa:2022tfv,Rosa:2023qcv}. Therefore, any matter accumulated within the star will {\it twinkle} for an external observer, making this a clear distinguishing feature of the oscillaton. In what follows, we analyze the emission profile of thin accretion disks in oscillatons.

\section{Accretion disks and the breathing emission}\label{sec:disks}
Let us now analyze the emission profile from an accretion disk around oscillatons. As discussed previously, since stable orbits exist in the entire extent of the radial coordinate, from the center of the oscillaton to infinity, we do not expect the presence of an ISCO cutoff, unlike in the Schwarzschild black hole case. This is so because oscillatons lack an event horizon and the strong-field potential barrier that such an ISCO typically produces; instead stable orbits are allowed to exist arbitrarily close to the center of the oscillaton until the influence of the scalar field starts to become non-negligible. Consequently, it is natural to anticipate that accretion may lead to the appearance of a central bright spot at the core of the star.  

The disk emission profile can be computed using several methods. One such method is using the Novikov–Thorne (NT) model~\cite{NovikovThorne,PageThorne}, where the radial profile of the emitted energy flux is given by  
\begin{equation}
    I(r)\propto-\frac{\Omega_{,r}}{\sqrt{-g}(E - \Omega L)^2}
    \int_{r_{\mathrm{in}}}^{r} (E - \Omega L)\, L_{,r'} \, dr',
    \label{eq:ntmodel}
\end{equation}
where the spacetime is assumed to be static, $E$ and $L$ denote the specific energy and specific angular momentum, respectively, $\Omega$ is the orbital frequency, and $r_{\rm in}$ is the inner edge of the disk. In the Schwarzschild case, a natural choice for the inner radius is the radius of the ISCO, while for dilute boson stars—where stable orbits exist for all $r$—one typically adopts $r_{\rm in}=0$.  

{We emphasize that the Novikov–Thorne model strictly applies to stationary spacetimes, where conserved quantities such as the specific energy are well defined. In the present case, the oscillaton spacetime is explicitly time dependent, and therefore the standard assumptions underlying the NT construction do not hold. For this reason, our use of NT-like profiles should be understood as a phenomenological emissivity ansatz, designed to isolate the effect of the time-dependent gravitational redshift on the observed intensity, rather than as a self-consistent accretion model. }

\begin{figure}[t!]
    \centering    \includegraphics[width=.9\linewidth]{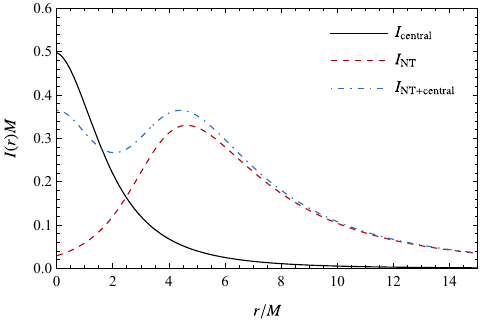}
    \caption{Accretion disk intensity profiles using the SU distribution of Eq. (\ref{eq:SU}). We consider three profiles, one that simulates a central emission ($I_{\rm central}$), one that fits a Novikov-Thorne profile ($I_{\rm NT}$), and one that combines both $(I_{\rm NT +  central})$.}
    \label{fig:profiles}
\end{figure}

{As mentioned above, the} NT model in Eq. \eqref{eq:ntmodel} strictly applies to stationary spacetimes, {we nevertheless adopt time-averaged orbital quantities,  as discussed in Sec.~\ref{sec:orbitstime}, as a proxy to construct illustrative emissivity profiles, which allow us to probe how the intrinsic time dependence of the geometry affects the observed emission. We also note that realistic accretion flows around supermassive black hole candidates such as Sgr A$^*$ and M87$^*$ are expected to be radiatively inefficient and geometrically thick, rather than thin, optically thick disks. While this may lead to quantitative differences in image morphology and intensity profiles, we do not expect it to alter the qualitative features discussed here.} This approximation allows us to estimate the emission intensity for oscillatons using Eq.~\eqref{eq:ntmodel}. For simplicity, we also fit the resulting intensity using the analytical family of models first considered in  \cite{Gralla:2020srx} and later generalized via suitable implementations of Johnson's standard unbound (SU)  distribution previously employed in the literature to reproduce specific scenarios of general-relativistic magnetohydroDynamic simulations of the accretion flow \cite{Gralla:2020srx,Vincent:2022fwj,Cardenas-Avendano:2023dzo},  as given by the three-parameter function
\begin{equation} \label{eq:SU}
    I(r)\propto\frac{\exp \left[-\frac{1}{2} \left(\gamma +\sinh ^{-1}\left(\frac{r-\beta }{\sigma }\right)\right)^2\right]}{\sqrt{(r-\beta )^2+\sigma ^2}}.
\end{equation}
This model conveniently accommodates both the central-emission configuration for the parameter choices \((\gamma,\beta,\sigma)=(0,0,2M)\), and the NT profile for \((\gamma,\beta,\sigma)=(-0.92,3.42M,2.46M)\). We also consider a model to accommodate both a central and the NT profile, formed by a composition of the two, scaled so that they have the same amplitude in the brightest spots. The corresponding intensity profiles as functions of the radius are shown in Fig.~\ref{fig:profiles}, where we see clear differences in the effective emission region of these central, NT, and central plus NT profiles.

\begin{figure*}
    \centering
    \includegraphics[width=0.33\linewidth]{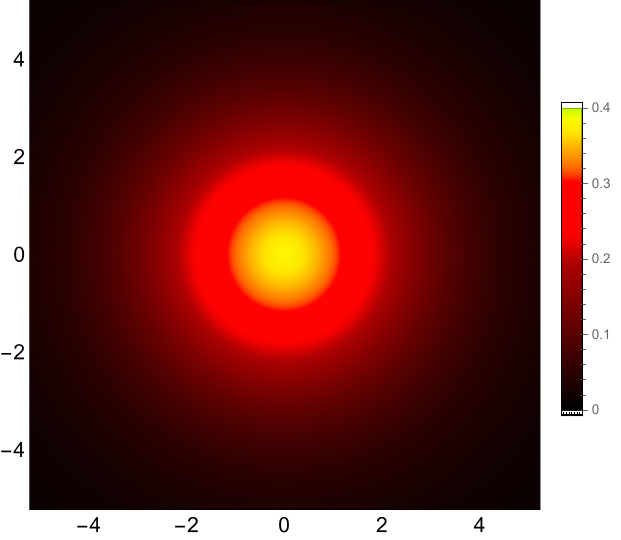}\includegraphics[width=0.33\linewidth]{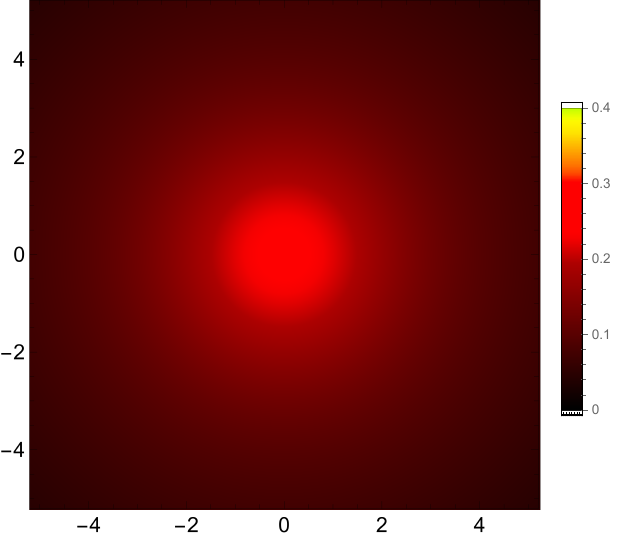}\includegraphics[width=0.33\linewidth]{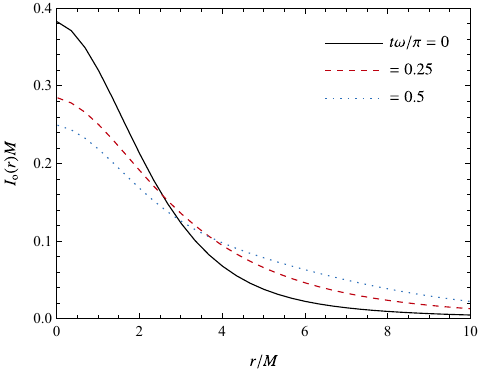}    \\
    \includegraphics[width=0.33\linewidth]{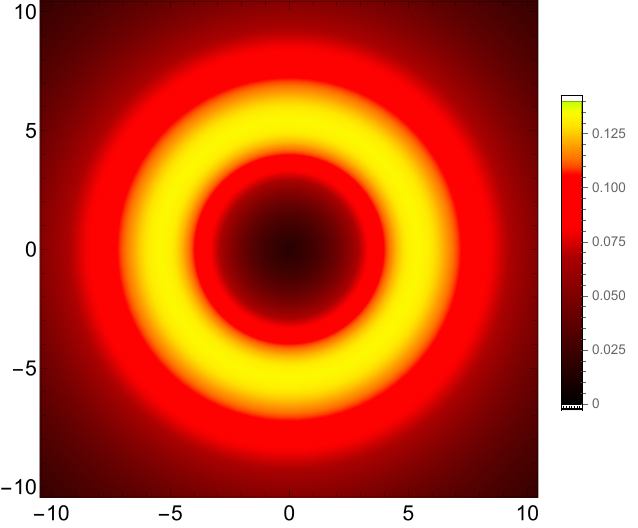}\includegraphics[width=0.33\linewidth]{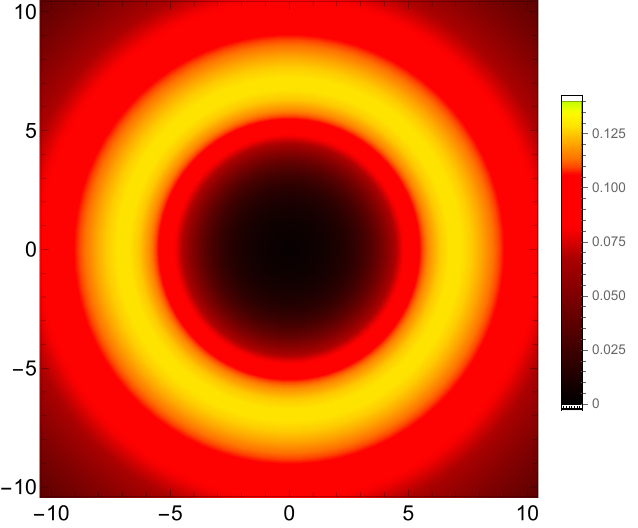}\includegraphics[width=0.33\linewidth]{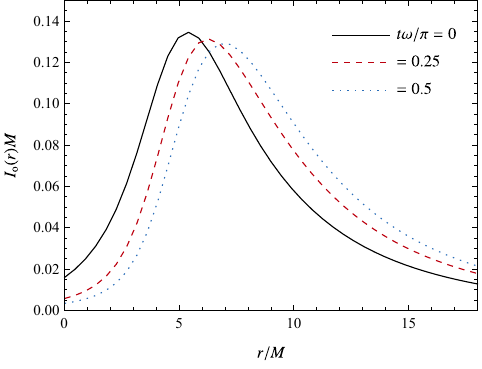}\\
    \includegraphics[width=0.33\linewidth]{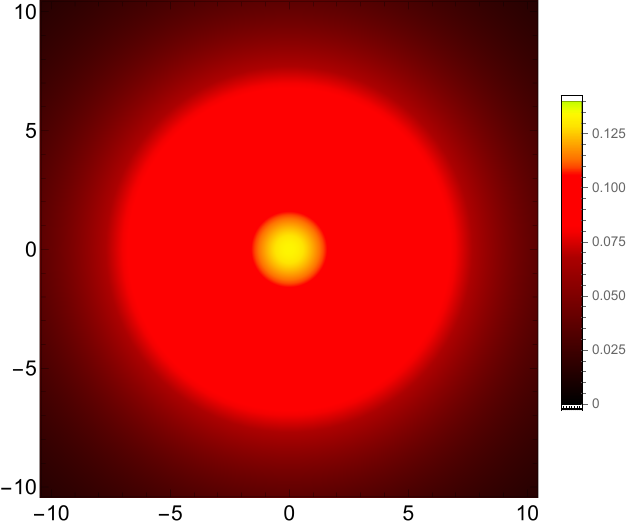}\includegraphics[width=0.33\linewidth]{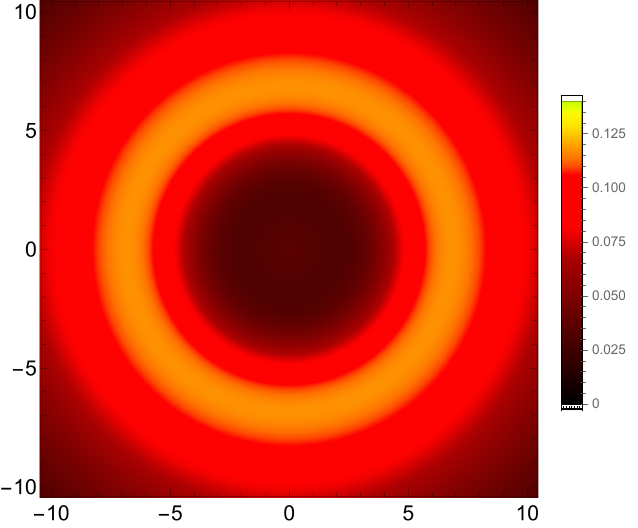}\includegraphics[width=0.33\linewidth]{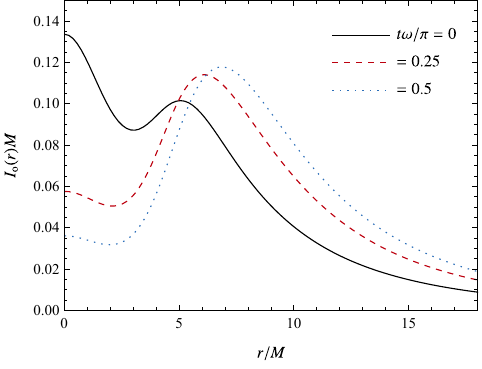}
    \caption{Intensity of the accretion disk profile observed at different instants of time. Due to the change in the redshift factor, the intensity changes with a period of $T=\pi/\omega$. The snapshots are given at $t=0$ (left column) and $t=\pi/(2\omega)$ (middle column). We show the result of a central intensity profile (upper row), the Novikov-Thorne-like emission profile (middle row), and a composition of both central and Novikov-Thorne-like profile (bottom row). We also show the observed intensity profiles for different values of time (third column).}
    \label{fig:axialshadow}
\end{figure*}

\begin{figure*}
\includegraphics[width=.33\linewidth]{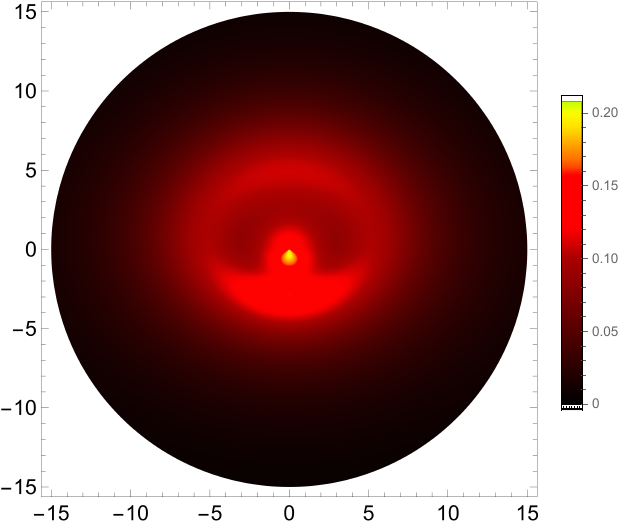}
\includegraphics[width=.33\linewidth]{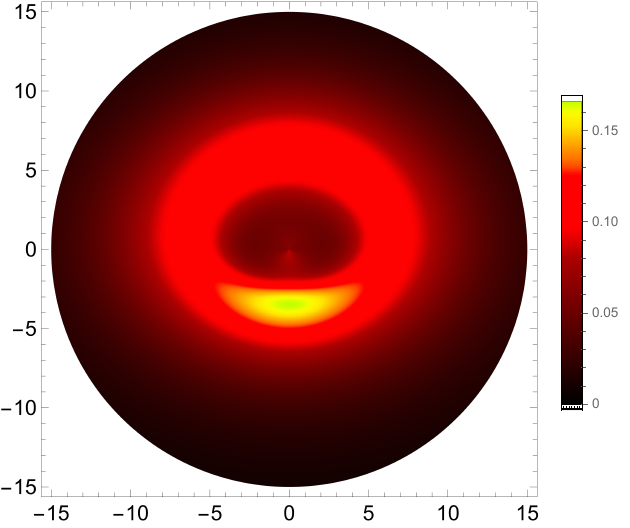}
\includegraphics[width=.33\linewidth]{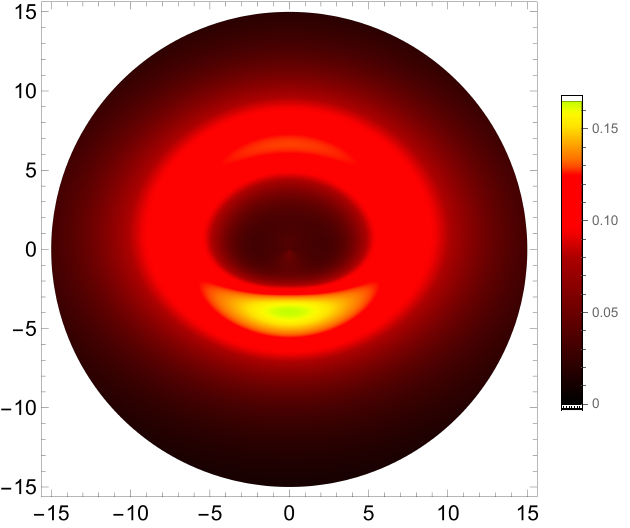}\\
\includegraphics[width=.33\linewidth]{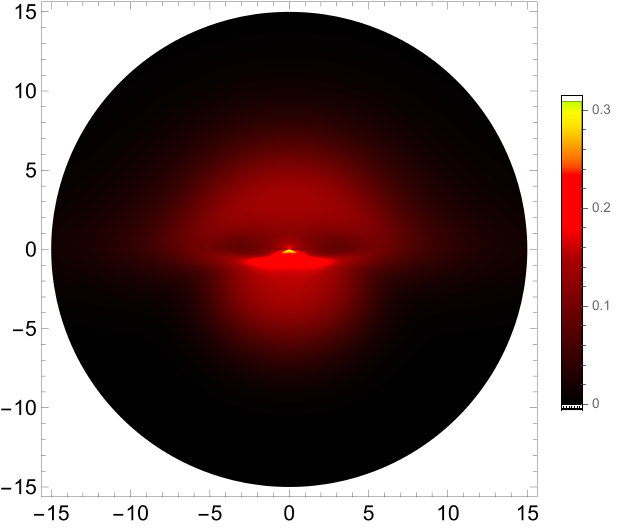}
\includegraphics[width=.33\linewidth]{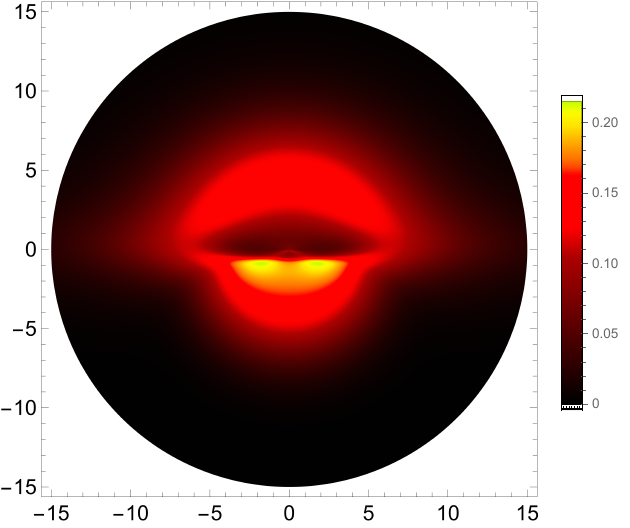}
\includegraphics[width=.33\linewidth]{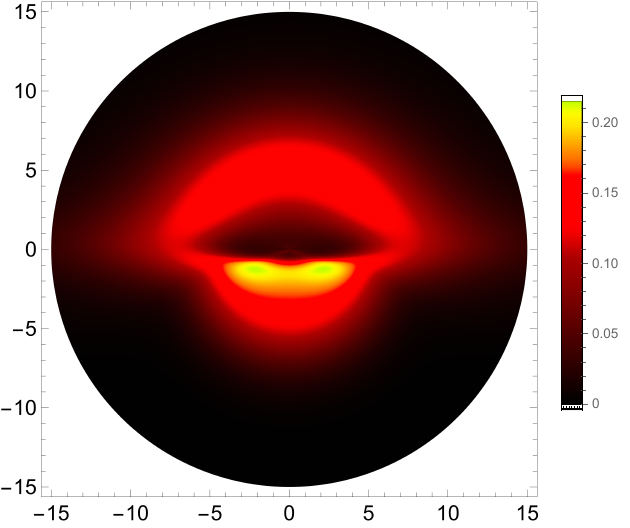}
    \caption{Intensity of the accretion disk with a central plus NT profile, observed at an angle of $\theta_{\rm o}=50^{\rm o}$ (top row) and  $80^{\rm o}$ (bottom row), considering $t\omega/\pi=0$ (left panel), 0.25 (middle panel), and 0.5 (right panel).}
    \label{fig:incshadow}
\end{figure*}

In Fig.~\ref{fig:axialshadow}, we show the axial observed images obtained from an accretion disk around oscillatons at two distinct instants of time, $t=0$ (left panels) and $t=\pi/(2\omega)$ (middle panels). The top panels correspond to the central emission case, while the middle panels depict the NT emission, and the bottom panels the central+NT profile. The temporal variation of the metric causes the emission profiles to change significantly, producing a characteristic ``breathing'' behavior as a function of time.  

iFor the central emission, the above feature of the oscillaton results in the central bright spot periodically fading in and out. No ``shadow" (i.e. a central brightness depression) is thus present in this model (as opposed to canonical black hole images), the oscillation appearing instead as a kind of luminous ``kink" reaching its maximum brightness at the center. In the NT case, we observe the presence of a central brightness depression enclosed by a luminous ring (similar to canonical black hole images), with the radius of the latter periodically expanding and contracting with time. Finally, for the central+NT profile there is a transition between a central luminous region and a brightness depression, which interchange as time passes by.

To further understand the origin of these images, as discussed previously, the observer perceives an intensity modulation due to the redshift between the source and the observer. The observed intensity $I_{\rm o}$ is therefore given by
\begin{equation}
    I_{\rm o}=A(t,r)^2 I_{\rm e}(r),
\end{equation}
as a consequence of the frequency redshift between emitter and observer and taking into account the shape of the line element in Eq. (\ref{eq:metric_g}), and where $I_{\rm e}$ denotes the intensity profile in the reference frame of the emitter. In the right panels of Fig.~\ref{fig:axialshadow} we display the observed intensity for the central, the NT, and the mixed NT + central model, where the temporal variations described above are evident. In the central model the observed intensity decreases monotonically from the center to asymptotic infinity for every oscillation, in the NT model the location of the peak of emission is slightly displaced on each oscillation, and for the combined central + NT model there are two peaks of observed intensity, whose relative heights are exchanged depending on the orbital period taken. Due to the absence of a horizon, it is clear that in these oscillation objects the presence (or not) of a shadow is determined by the amount of residual luminosity present at the center (measured in terms of the fraction of the peak luminosity), meaning that oscillaton objects might pass EHT observations depending on the actual emission features of the accretion disk.

Regarding the latter point, it is worth assessing whether the oscillations described above could be resolved in observations of accretion disk profiles around supermassive objects. Restoring physical units, the period of the background field is
\begin{equation}
    T = (\omega/\pi)^{-1} \sim 28.55 \left( \frac{M}{10^6\, M_\odot} \right) {\rm s}.
\end{equation}
For the two primary EHT targets, this corresponds to $T \sim 2\,{\rm min}$ for Sgr~A$^{*}$ and $T \sim 19\,{\rm h}$ for M87$^{*}$. These timescales suggest that, in principle, the EHT could detect such variability within the duration of its observing campaigns~\cite{EventHorizonTelescope:2019dse,EventHorizonTelescope:2022wkp}. {We emphasize that these images are idealized and assume an angular resolution superior to that currently achievable with the EHT. This issue is examined in more detail in Appendix~\ref{ap:blurring}.} Consequently, if supermassive oscillatons reside at the centers of galaxies, the observation of these twinkling features would provide compelling evidence for their existence. Moreover, since scalar fields can naturally accumulate around a variety of compact objects---including rotating black holes---similar oscillatory signatures may arise for \emph{any} supermassive central object, in addition to other observational imprints~\cite{Ferreira:2017pth}. We note, however, that the oscillation period depends on the specific configuration and may vary across the parameter space.

Finally, for completeness, in Fig.~\ref{fig:incshadow} we show the intensity profiles for inclined observations at $\theta_{\rm o}=50^{\circ}$ (upper panels) and $\theta_{\rm o}=80^{\circ}$ (bottom panels). We display snapshots (from left to right) at $t=0$,  $t=\pi/(4\omega)$ and $t=\pi/(2\omega)$. We focus on the case with a central plus NT emission profile, which exhibits the most distinctive behavior. The resulting images share several features with the axial view, including the periodic transition from a centrally bright region to ringlike structures as the oscillation progresses. Inclination, however, brings out an additional characteristic: gravitational lensing. It is remarkable that oscillatons, despite not being sufficiently compact to support light rings, still display strong qualitative similarities to black hole images---though with important differences, such as the absence of photon-ring structures.

\begin{figure}
    \centering
    \includegraphics[width=1\linewidth]{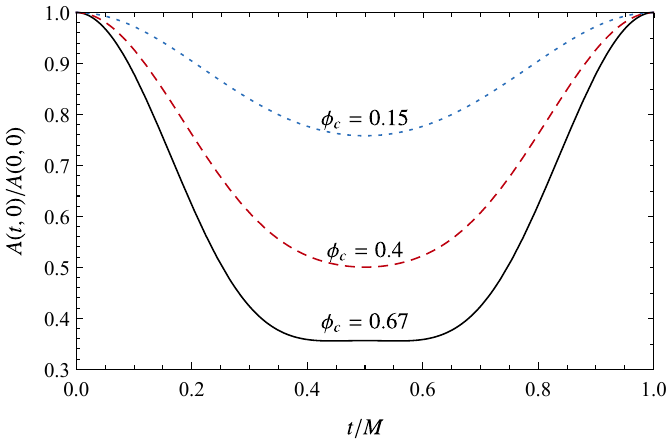}
    \caption{{Redshift factor at the center of the oscillaton for three different configurations, given by $(\phi_c,\omega/\mu,M\mu)=(0.67,0.864,0.604)$, $(0.4,0.912,0.571)$, and $(0.15,0.965,0.423)$. For less compact configurations, the relative variation of the redshift factor between its minimum and maximum values decreases, indicating that the ``twinkling'' effect becomes weaker.}}
    \label{fig:redshift}
\end{figure}

{One might wonder whether less compact configurations can produce similar modulations in the disk intensity profile. A simple way to assess this is to examine the redshift factor $A(t,r)$, particularly near the origin, where the effect is strongest. Since the intensity modulation depends intrinsically on how the redshift changes as the metric evolves, in Fig.~\ref{fig:redshift} we show the redshift factor at the center, normalized by its value at $t=0$. The configurations shown are characterized by $(\phi_c,\omega/\mu,M\mu)=(0.67,0.864,0.604)$, $(0.4,0.912,0.571)$, and $(0.15,0.965,0.423)$. Even the least compact configuration, with $\phi_c=0.15$ and effective radius $R\approx 42.95M$, still exhibits a difference of about $20\%$ between the minimum and maximum values of the redshift factor. This indicates that less compact configurations can still produce a considerable dimming of the brightness. This would be particularly important for black holes immersed in lesser dense scalar media~\cite{Ferreira:2017pth}, however one still needs to check how the BH changes the brightness modulation.}

\section{Conclusion}  \label{sec:conclusion}

In this work, we have explored the observational phenomenology of oscillatons, a class of
time-periodic solitonic configurations supported by real scalar fields. By constructing
fully relativistic oscillaton solutions through a Fourier-expanded decomposition of the
Einstein--Klein--Gordon equations, we identified the most compact configurations and
characterized their spacetime structure. We demonstrated that these objects support
oscillatory circular orbits at all radii, including regions well inside the stellar interior, and that slight perturbations of these orbits exhibit epicyclic modulations that closely
track the predictions obtained from the time-averaged metric. These results extend the
parallel between oscillatons and boson stars, highlighting that stable bound orbits are a
robust feature of light-field solitons.

We further analyzed light propagation in oscillaton geometries, revealing the presence of
caustics, rainbow scattering, and partial visibility of emitters located inside the star.
These lensing features, while qualitatively similar to those of fluid stars or boson stars,
acquire an additional layer of complexity due to the underlying time dependence of the
metric. This time dependence has particularly striking consequences for accretion-disk
imaging: the redshift oscillations modulate the observed intensity, producing a
characteristic breathing pattern in which central bright regions and ringlike structures
alternate over a timescale set by the scalar-field mass. Using central-emission, 
Novikov--Thorne-like models, and a combination of both, we showed that this variability is robust across different
accretion prescriptions and persists even for inclined viewing angles.

Finally, we assessed the relevance of these features for supermassive oscillatons that
might reside at galactic centers. Restoring physical units, we found that the oscillation
period lies within the observational window of the Event Horizon Telescope for the
masses appropriate to Sgr~A$^{*}$ and M87$^{*}$. This suggests that horizon-scale imaging may
be able to test, or potentially constrain, the existence of oscillatonlike configurations
through the detection (or absence) of such ``twinkling'' behavior. More broadly, our
results reinforce the idea that time-dependent solitonic structures sourced by ultralight
fields can imprint distinctive signatures on electromagnetic observables, providing a
promising avenue for probing beyond-Standard-Model physics in the strong-gravity
regime.

Future work may include extending these results to rotating oscillatons, incorporating
radiative transfer in more realistic accretion flows, and examining the interplay between
oscillaton dynamics and the evolution of surrounding matter fields. {The results presented here should be interpreted as a proof of principle: we demonstrate that time-dependent spacetimes can imprint periodic signatures on observables. A full assessment of detectability, including realistic accretion physics and interferometric reconstruction, is beyond the scope of this work.} These developments
would further clarify the astrophysical viability of oscillatons and sharpen their
observational predictions. Work along these lines is currently underway.

\section*{Acknowledgements}
This work is supported by the Spanish National Grants PID2022-138607NBI00 and CNS2024-154444, funded by MICIU/AEI/10.13039/501100011033 (“PGC Generación de Conocimiento") and FEDER, UE. C.F.B.M. acknowledges Coordenação de Aperfeiçoamento de Pessoal de Nível Superior– Brasil (CAPES)– Finance Code 001, Fundação Amazônia de Amparo a Estudos e Pesquisa (FAPESPA), and Conselho Nacional de Desenvolvimento Científico e Tecnológico (CNPq).

\appendix
\section{EHT observation of oscillating boson stars} \label{ap:blurring}
\begin{figure*}
\includegraphics[width=0.8\linewidth]{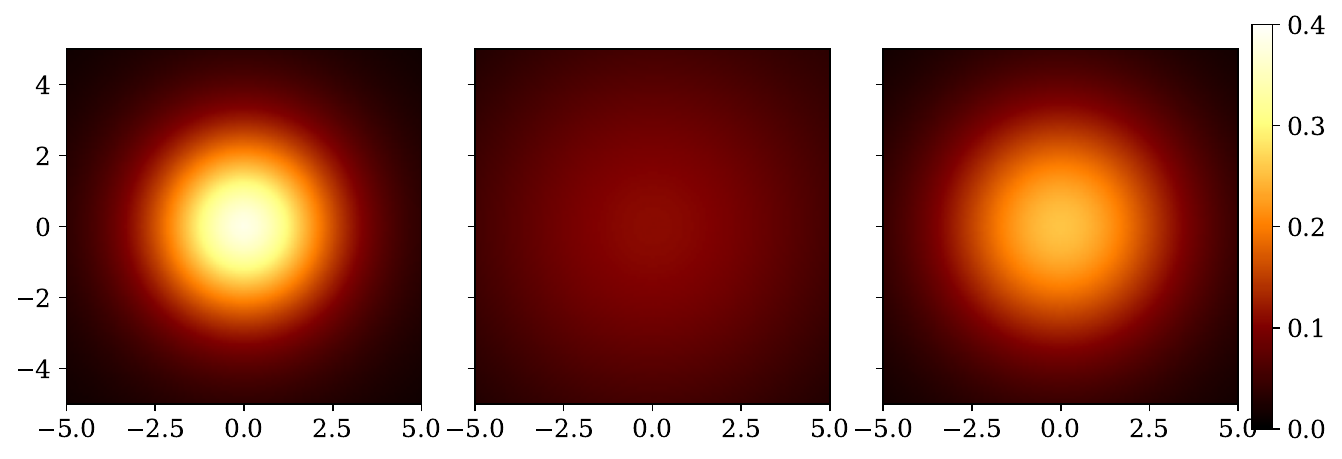}\\
\includegraphics[width=0.8\linewidth]{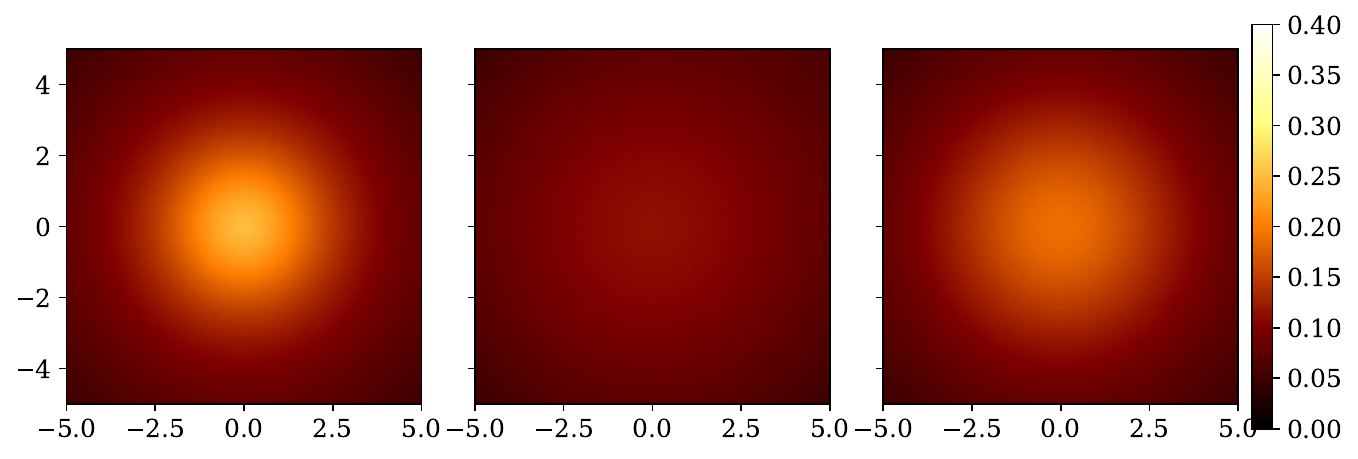}
    \caption{Intensity maps of the accretion disk with a central profile (face-on): unblurred (left), convolved with a $\sim 15 \, \mu as$ beam (middle) representing the EHT resolution, and with a $\sim 6 \, \mu as$ beam (right) representing the anticipated EHT+BHEX resolution. The snapshots are given at $t=0$ (first row) and $t=\pi/(2\omega)$ (second row).}
    \label{fig:blur}
\end{figure*}

\begin{figure*}
\includegraphics[width=0.8\linewidth]{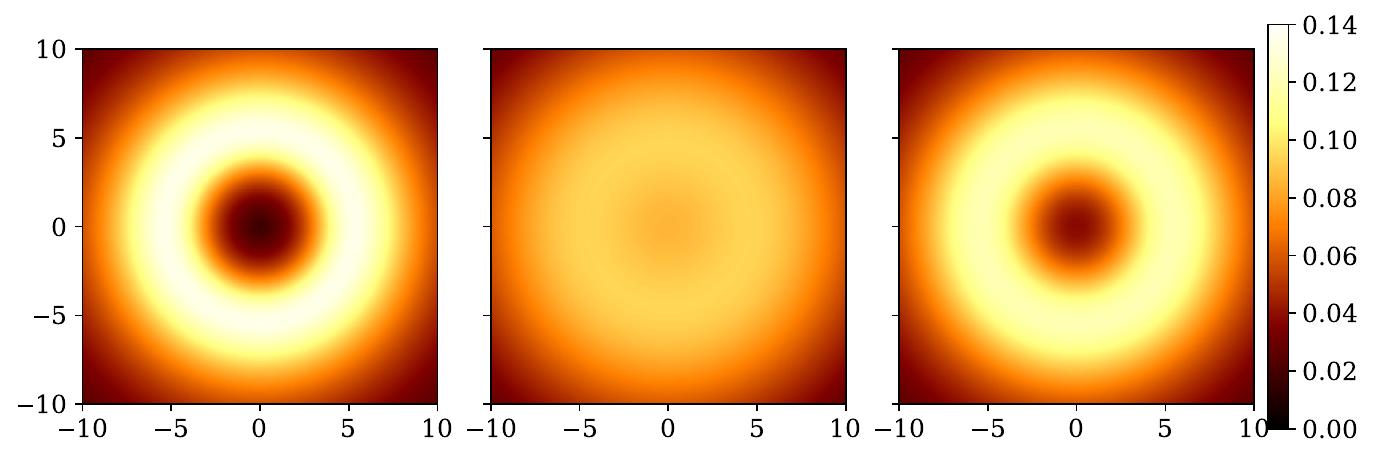}\\
\includegraphics[width=0.8\linewidth]{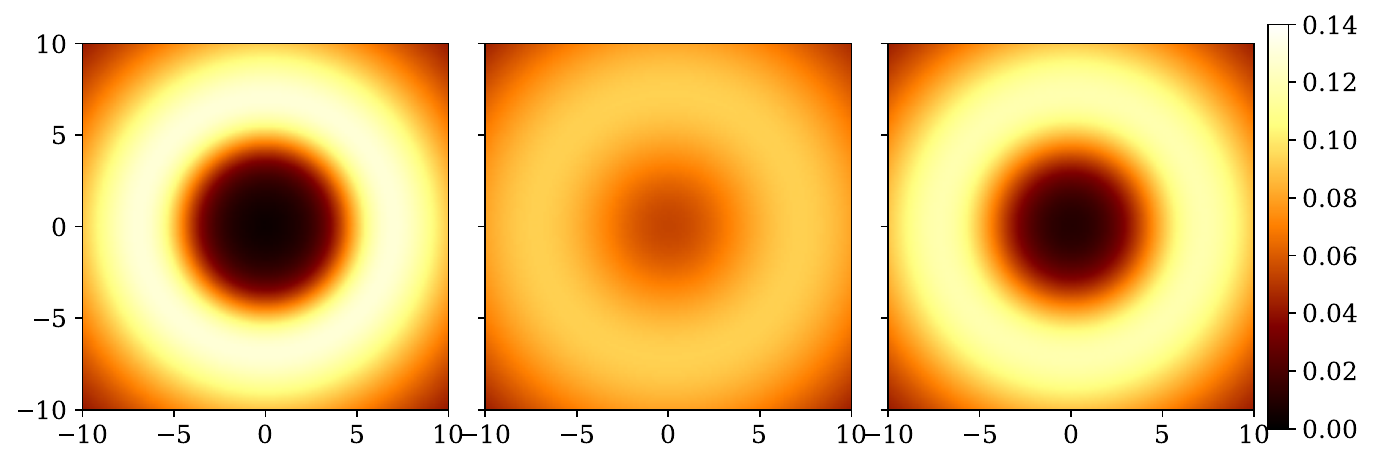}
    \caption{Intensity maps of the accretion disk with a NT profile (face-on): unblurred (left), convolved with a $\sim 15 \, \mu as$ beam (middle) representing the EHT resolution, and with a $\sim 6 \, \mu as$ beam (right) representing the anticipated EHT+BHEX resolution. The snapshots are given at $t=0$ (first row) and $t=\pi/(2\omega)$ (second row).}
    \label{fig:blur2}
\end{figure*}

\begin{figure*}
\includegraphics[width=0.7\linewidth]{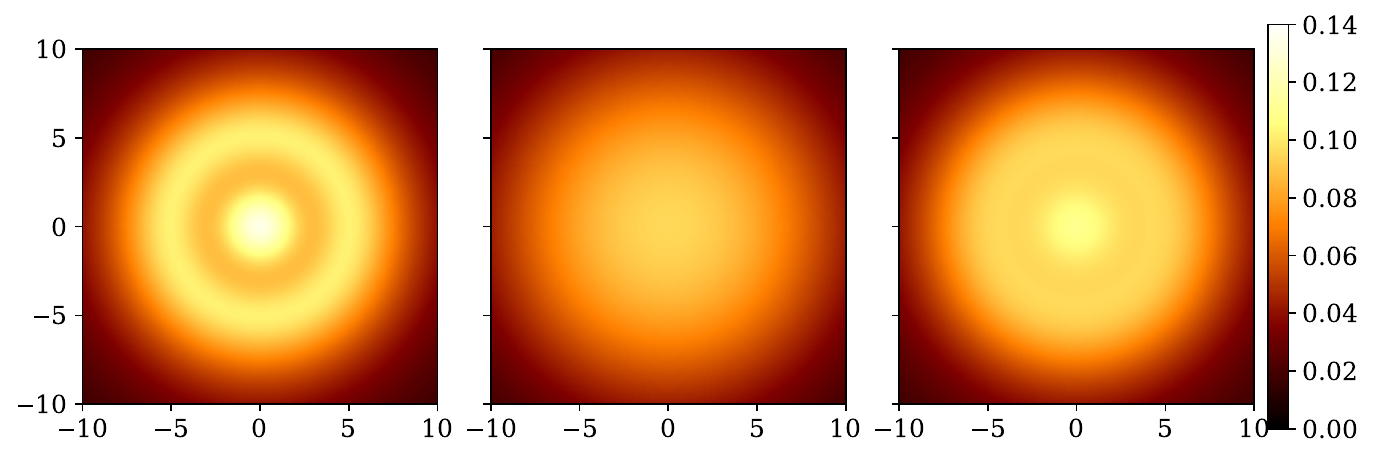}\\
\includegraphics[width=0.7\linewidth]{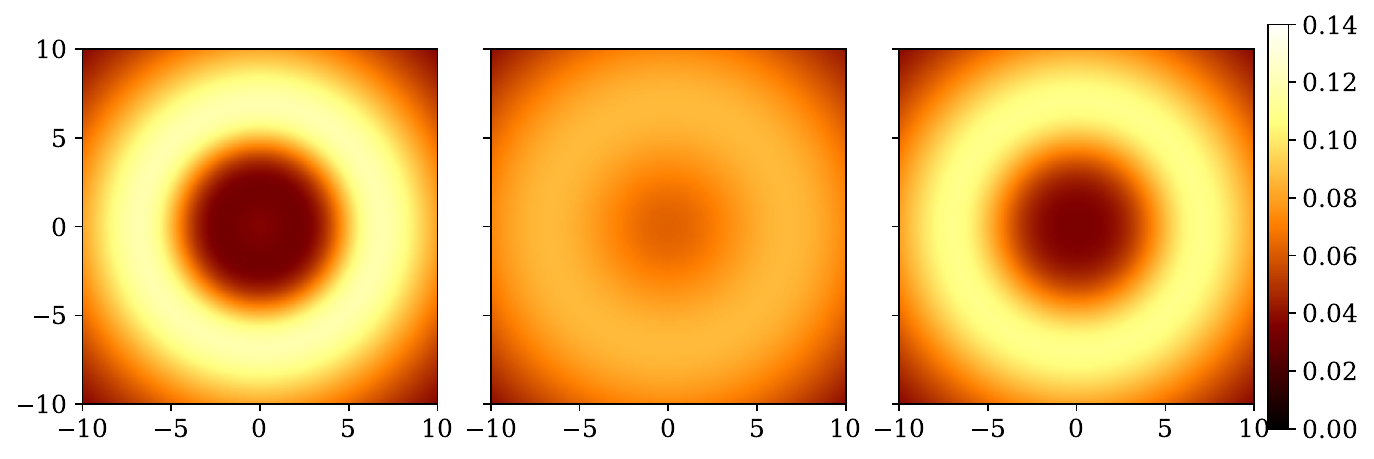}
    \caption{Intensity maps of the accretion disk with a central plus NT profile (face-on): unblurred (left), convolved with a $\sim 15 \, \mu as$ beam (middle) representing the EHT resolution, and with a $\sim 6 \, \mu as$ beam (right) representing the anticipated EHT+BHEX resolution. The snapshots are given at $t=0$ (first row) and $t=\pi/(2\omega)$ (second row).}
    \label{fig:blur3}
\end{figure*}

\begin{figure*}
\includegraphics[width=0.7\linewidth]{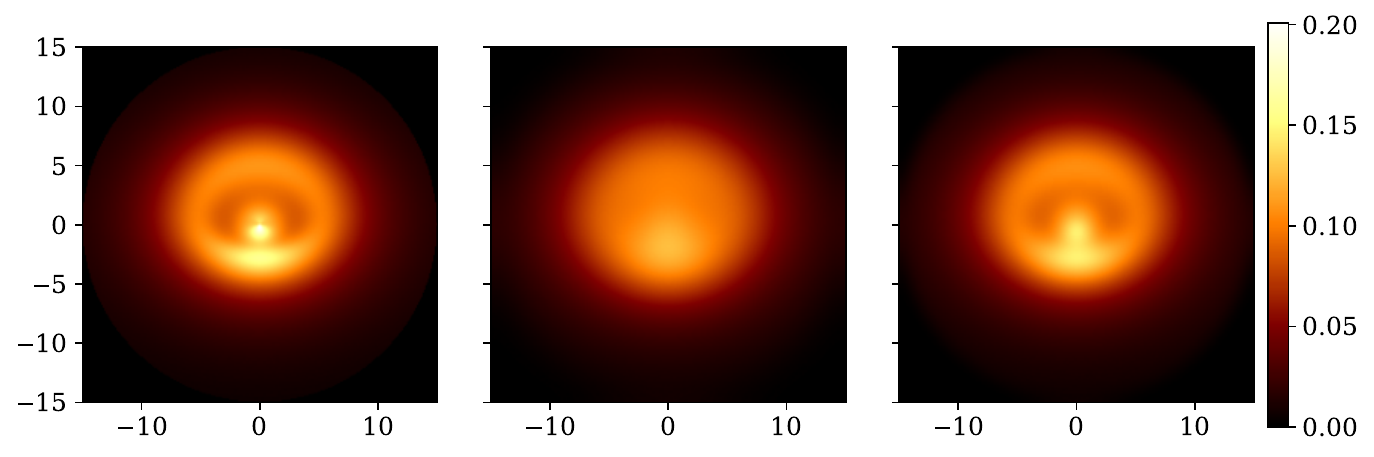}\\
\includegraphics[width=0.7\linewidth]{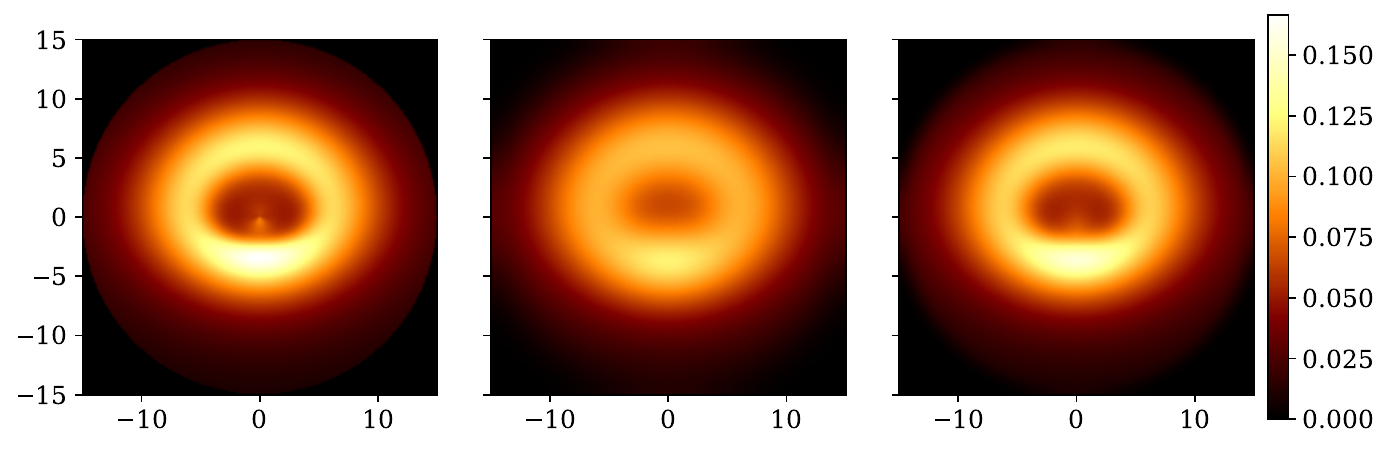}\\
\includegraphics[width=0.7\linewidth]{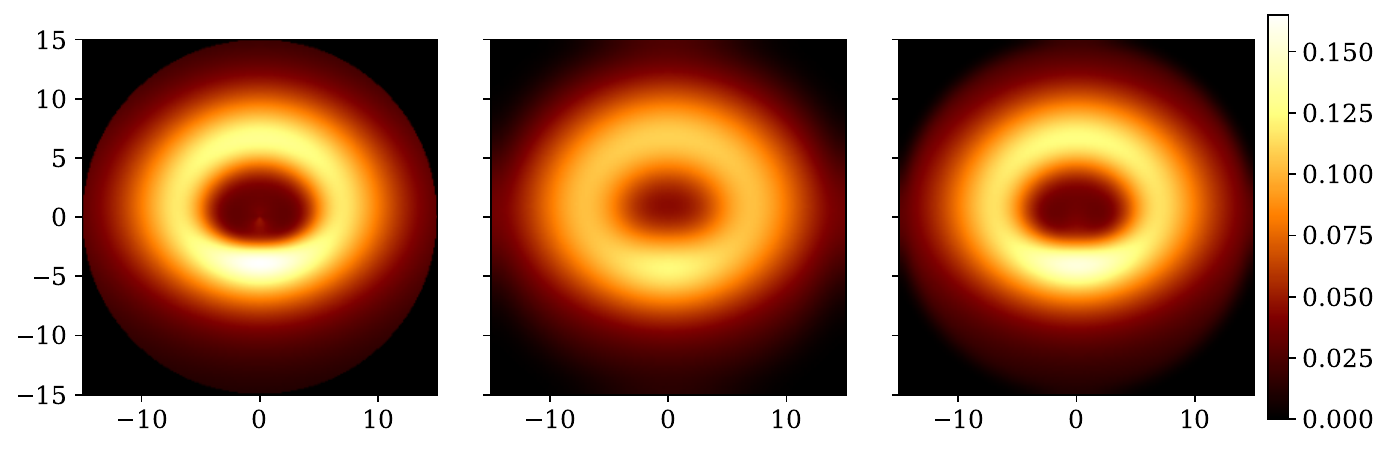}
    \caption{Intensity maps of the accretion disk with a central profile plus NT profile with a $50 ^\circ$ inclination: unblurred (left), convolved with a $\sim 15 \, \mu as$ beam (middle) representing the EHT resolution, and with a $\sim 6 \, \mu as$ beam (right) representing the anticipated EHT+BHEX resolution. The snapshots are given at $t \omega / \pi=0$ (first row), $t \omega / \pi= 0.25$ (second row) and $t \omega / \pi= 0.5$ (third row).}
    \label{fig:blur4}
\end{figure*}

In this section, we examine how these objects would appear from an observational perspective, first at the angular resolution currently achievable by the EHT, approximately $\sim 15 \, \mu\mathrm{as}$, and then at the improved resolution anticipated from the combined capabilities of the EHT and the forthcoming Black Hole Explorer (BHEX) mission \cite{Johnson:2024ttr}, which is expected to reach $\sim 6 \, \mu\mathrm{as}$. To mimic the effect of finite instrumental resolution, we convolve the synthetic images with a Gaussian kernel whose full width at half maximum (FWHM) matches the chosen angular resolution. The correspondingly blurred images for the different configurations discussed above are presented in Figs.~\ref{fig:blur}–\ref{fig:blur4}, where we display them together with the original, unblurred images for direct comparison.

For the face-on central-disk profile shown in Fig.~\ref{fig:blur}, the change in intensity produced by variations in the Doppler factor due to metric oscillations in the central region between $t=0$ (top row) and $t=\pi/(2\omega)$ (bottom row) cannot be resolved with the current EHT angular resolution, as the blurring smooths out the central brightness differences. In contrast, this intensity modulation becomes visible when the higher resolution of EHT+BHEX is used. We further stress that none of these images are consistent with present observations, since they do not exhibit a central shadow, which necessarily follows from the assumed disk profile.
When we switch to an NT-disk configuration (Fig.~\ref{fig:blur2}), a central shadow does appear. The associated variation in the apparent shadow size as the metric oscillates is, however, challenging to detect with the current resolving power of the EHT. This variation should, on the other hand, be straightforward to measure with the enhanced angular resolution of the combined EHT+BHEX array.
Lastly, for the central-plus-NT disk profile (Fig.~\ref{fig:blur3}), the EHT-resolution images already show substantial differences between the two oscillation phases, making the variability clearly visible. These differences become even more pronounced when the combined EHT+BHEX angular resolution is employed. For the $50^\circ$ inclination scenario with the central-plus-NT disk profile shown in Fig.~\ref{fig:blur4}, the above conclusions still hold. In addition, the blurred images at EHT resolution closely resemble the M87* observations, particularly at $t \omega / \pi = 0.25$ and $t \omega / \pi = 0.5$.

\bibliographystyle{unsrt}
\bibliography{main}

\end{document}